\documentclass[12pt]{article}

\usepackage{epsf,amsfonts,amssymb,epsfig,amsmath}

\renewcommand{\text}[1]{#1}

\newcommand{\be}{\begin{equation}}
\newcommand{\ee}{\end{equation}}
\newcommand{\ben}{\begin{displaymath}}
\newcommand{\een}{\end{displaymath}}
\newcommand{\bea}{\begin{eqnarray}}
\newcommand{\eea}{\end{eqnarray}}
\newcommand{\bean}{\begin{eqnarray*}}
\newcommand{\eean}{\end{eqnarray*}}
\newcommand{\nn}{\nonumber \\}
\newcommand{\ba}{\begin{array}}
\newcommand{\ea}{\end{array}}
\newcommand{\bi}{\begin{itemize}}
\newcommand{\ei}{\end{itemize}}

\newcommand{\reef}[1]{(\ref{#1})}

\def\G{\Gamma}

\def\e{\epsilon}

\def\otaula{\begin{tabular}}
\def\ctaula{\end{tabular}}

\def\bnum{\begin{enumerate}}
\def\enum{\end{enumerate}}

\def\CR{\mathcal{R}}
\def\CM{\mathcal{M}}

\def\8M{$\CM_8$}

\def\be{\begin{equation}}
\def\ee{\end{equation}}
\def\G{\Gamma}

\def\ei{e^{\underline{i}}}

\def\e1{e^{\underline{1}}}
\def\1u{\underline{1}}
\def\2u{\underline{2}}

\def\0u{\underline{0}}
\def\e{\epsilon}
\def\target{$\CR^{1,1}\times \mathcal{M}_8$ }
\def\target2{$\CR^{1,1}\times \mathcal{M}_8$,}
\def\9G{\G_{\underline{9}}}

\newcommand{\bbR}{{\mathbb{R}}}

\newcommand{\bbC}{{\mathbb{C}}}

\newcommand{\CP}{\mathbb{C}P}

\DeclareMathOperator{\vol}{vol}

\newcommand{\del}{\partial}

\newcommand{\zG}{{z}}

\def\1f{f_1^{1/2}}
\def\2f{f_2^{1/2}}
\def\4f{f_4^{1/2}}

\begin{document}

\makeatletter
\renewcommand{\theequation}{\thesection.\arabic{equation}}
\@addtoreset{equation}{section}
\makeatother

\baselineskip 18pt

\begin{titlepage}

\vfill

\begin{flushright}
Imperial/TP/2006/JG/03\\
hep-th/0612253\\
\end{flushright}

\vfill

\begin{center}
   \baselineskip=16pt
   {\Large\bf  Supersymmetric $AdS_3$, $AdS_2$ and Bubble Solutions}
   \vskip 2cm
      Jerome P. Gauntlett$^1$, Nakwoo Kim$^2$
      and Daniel Waldram$^1$
   \vskip .6cm
      \begin{small}
      $^1$\textit{Theoretical Physics Group, Blackett Laboratory, \\
        Imperial College, London SW7 2AZ, U.K.}
        \end{small}\\*[.6cm]
      \begin{small}
      $^1$\textit{The Institute for Mathematical Sciences, \\
        Imperial College, London SW7 2PE, U.K.}
        \end{small}\\*[.6cm]
      \begin{small}
      $^2$\textit{Department of Physics and Research Institute of
        Basic Science, \\
        Kyung Hee University, Seoul 130-701, Korea}
        \end{small}
   \end{center}

\vfill

\begin{center}
\textbf{Abstract}
\end{center}
We present new supersymmetric $AdS_3$ solutions of type IIB supergravity and
$AdS_2$ solutions of $D=11$ supergravity. The former are dual to
conformal field theories in two dimensions with $N=(0,2)$
supersymmetry while the latter are dual to conformal quantum mechanics
with two supercharges. Our construction also includes $AdS_2$
solutions of $D=11$ supergravity that have non-compact internal spaces
which are dual to three-dimensional $N=2$ superconformal field
theories coupled to point-like defects. We also present some new
bubble-type solutions, corresponding to BPS states in conformal
theories, that preserve four supersymmetries.

\vfill

\end{titlepage}
\setcounter{equation}{0}


\section{Introduction}

Supersymmetric solutions of $D=10$ and $D=11$ supergravity that
contain $AdS$ factors are dual to superconformal field theories
(SCFTs). It is therefore of interest to study the generic geometric
structure of such solutions and, in particular, to use this insight to
construct new explicit solutions.

The most general supersymmetric solutions of type IIB supergravity
with an $AdS_3$ factor and with only the five-form flux non-trivial
were analysed in~\cite{nak1}. These solutions can arise as the
near-horizon geometry of configurations of D3-branes, preserve 1/8-th
of the supersymmetry and are dual to two-dimensional $N=(0,2)$
superconformal field theories. Similarly, in~\cite{nak2} the most
general supersymmetric solutions of $D=11$ supergravity with an
$AdS_2$ factor and a purely electric four-form flux were
analysed. These solutions can arise as the near-horizon geometry of
configurations of M2-branes, also preserve 1/8-th of the supersymmetry
and are dual to superconformal quantum mechanics with two supercharges.

We shall summarise the main results of~\cite{nak1,nak2} in section 2
below. What is remarkable is that the internal manifolds in each case
have the same geometrical structure. For the type IIB $AdS_3$
solutions, locally the seven-dimensional internal manifold $Y_7$ has a
natural foliation, such that the metric is completely determined by a
K\"ahler metric on the six-dimensional leaves. For the $D=11$ $AdS_2$
solutions, locally the metric on the internal manifold $Y_9$ is again
completely determined by a K\"ahler metric on, now, eight-dimensional
leaves. Both $(2n+2)$-dimensional K\"ahler metrics $ds^2_{2n+2}$
satisfy exactly the same differential condition
\be
\Box R -\tfrac{1}{2} R^2 + R_{ij} R^{ij} = 0
\label{mm}
\ee
where $R$ and $R_{ij}$ are the Ricci-scalar and Ricci-tensor,
respectively, of the metric $ds^2_{2n+2}$. In each case, to obtain
an $AdS_3$ or $AdS_2$ solution one requires $R>0$.

It is worth noting the similarities with Sasaki-Einstein (SE) metrics.
Recall that five-dimensional SE metrics, $SE_5$ give rise to supersymmetric
solutions of type IIB supergravity of the form $AdS_5\times SE_5$,
while seven-dimensional SE metrics, $SE_7$ give rise to supersymmetric
solutions of $D=11$ supergravity of the form $AdS_4\times SE_7$.
All SE metrics have a canonical Killing vector which defines, at least locally,
a canonical foliation, and the SE metric is completely determined
by a K\"ahler-Einstein metric on the corresponding leaves.
There has been some recent explicit constructions of local K\"ahler-Einstein metrics
that give rise to complete SE metrics and we will show that they
can be adapted to produce K\"ahler metrics that satisfy \reef{mm} and hence
give rise to new $AdS_3$ and $AdS_2$ solutions.

After an analytic continuation, the generic $AdS_3$ and $AdS_2$
solutions discussed in~\cite{nak1,nak2} give rise to generic
supersymmetric solutions with $S^3$ and $S^2$ factors preserving
1/8-th of  maximal supersymmetry. In particular the solutions are built
from the same K\"ahler geometry satisfying~\reef{mm}, but now with $R<0$.
Such ``bubble solutions'' generalise the 1/2 supersymmetric bubble
solutions of~\cite{llm} (1/4 supersymmetric bubble solutions in type IIB were analysed
in~\cite{donos1,donos2}) and generically have an $\bbR\times SO(4)$
or $\bbR\times SO(3)$ group of isometries.  Depending on the boundary
conditions, the 1/8-th supersymmetric bubbles can describe 1/8-th BPS states
in the maximally SCFTs, or other BPS states in SCFTs with less
supersymmetry. Note that recently an analysis of 1/8-th supersymmetric
bubbles in type IIB supergravity with additional symmetries was
carried out in~\cite{gava} and, most recently, $AdS_2$ and bubble
solutions of $D=11$ supergravity preserving various amounts of
supersymmetry were analysed in \cite{eo}. The constructions that we
use for the $AdS_3$ and $AdS_2$ solutions, that we outline below, also
lead to new explicit bubble solutions.

We will present three constructions of K\"ahler metrics
satisfying~\reef{mm} which lead to new $AdS$ and bubble solutions. The first
construction is directly inspired by the constructions of SE metrics
in~\cite{Gauntlett:2004yd,Gauntlett:2004hh}. Following~\cite{Page:1985bq},
the idea is to build the local K\"ahler metric from a $2n$-dimensional K\"ahler--Einstein
metric $ds^2(KE_{2n})$. To construct complete metrics on the internal
space $Y_{2n+3}$ we will usually assume that the $2n$-dimensional
leaves on which $ds^2(KE_{2n})$ is defined extend to form a compact
K\"ahler--Einstein space $KE_{2n}$. One might then try to similarly
extend the K\"ahler metrics $ds^2_{2n+2}$ to give non-singular metrics
on a compact space which is an $S^2$ fibration over $KE_{2n}$. However, this
is not possible. Nonetheless, as we show, this kind of construction can give rise
to complete and compact metrics on $Y_{2n+3}$. This is precisely
analogous to the construction of Sasaki--Einstein manifolds presented
in~\cite{Gauntlett:2004yd,Gauntlett:2004hh}. For the six-dimensional
case, we show that we recover the $AdS_3$ solutions of type IIB
supergravity that were recently constructed in~\cite{gmmw2}. On the
other hand, the eight-dimensional case leads to new infinite classes
of $AdS_2$ solutions of $D=11$ supergravity.

In~\cite{gmmw2} it was shown that by choosing the range of coordinates
differently, one obtains $AdS_3$ solutions with non-compact internal
spaces. These solutions were interpreted as being dual to
four-dimensional ${\cal N}=1$ SCFTs, arising from five-dimensional
Sasaki-Einstein spaces, in the presence of a one-dimensional
defect. Similarly, we can also choose the range of the coordinates in
the new $AdS_2$ solutions presented here so that they also have
non-compact internal spaces. As we discuss, these solutions are dual
to three-dimensional ${\cal N}=2$ SCFTs arising from seven-dimensional
Sasaki-Einstein spaces, in the presence of a point-like defect.

We will then show that this first construction of K\"ahler metrics
also gives rise to supersymmetric bubble solutions. Indeed,
remarkably, we find that we recover the uplifted versions of the
$AdS$ single-charged ``black hole'' solutions of minimal gauged
supergravity in $D=5$ and $D=4$. Recall that these BPS solutions
have naked singularities and hence were christened ``superstars'' in
\cite{Myers:2001aq,Leblond:2001gn}. These solutions are special
cases of more general superstar solutions obtained by uplifting
three- and four-charged $AdS$ ``black holes'' in $D=5$ and $D=4$
gauged supergravity, respectively. We identify the underlying
K\"ahler geometry for these general solutions which we then employ
to carry out our second construction of supersymmetric $AdS3$ and
$AdS_2$ solutions. The K\"ahler geometries are toric and hence this
construction is analogous to the construction of SE metrics of
\cite{Cvetic:2005ft} (see also \cite{Martelli:2005wy}).

The third construction of K\"ahler metrics satisfying~\reef{mm}
that we shall present is to simply take the metric to be a direct
product of K\"ahler-Einstein metrics. This gives rise to rich new
classes of $AdS_3$ and $AdS_2$ solutions. A special case of the
$AdS_3$ solutions is that given in~\cite{naka}, describing
$D3$-branes wrapping a holomorphic curve in a Calabi-Yau four-fold,
while a special case of the $AdS_2$ solutions corresponds to the solution
in~\cite{gkwp}, describing $M2$-branes wrapping a holomorphic curve
in a Calabi-Yau five-fold. The construction also gives rise to infinite new
bubble solutions.

The plan of the rest of the paper is as follows. We begin in section 2 by
reviewing the construction of \cite{nak1,nak2}. In section 3 we describe
the construction using $S^2$ fibrations over $KE$ manifolds.
In section 4 we show that this construction gives rise to bubble solutions that
are the same as the uplifted single charged superstars.
In section 5, we determine the K\"ahler geometry underlying the multiple charged
superstars and then use this to construct infinite new class
of $AdS$ solutions. In section 6 we describe the construction of $AdS$ and
bubble solutions using products of KE metrics. Section 7 briefly concludes.


\section{Background}

\subsection{$AdS_3$ in IIB and $AdS_2$ in $D=11$}

The generic $AdS_3$ and $AdS_2$ solutions discussed
in~\cite{nak1,nak2} are constructed as follows. In each case, one
assumes the metric is a warped product
\begin{equation}
\label{firsteq}
   ds^2 = L^2 e^{2A} \left[ ds^2(AdS_d) + ds^2(Y_{2n+3}) \right]
\end{equation}
where we normalise such that $ds^2(AdS_d)$
has unit radius and $L$ is an overall scale factor that we will
sometimes normalise to one.
Let $ds^2_{2n+2}$ be a $2n+2$-dimensional K\"ahler metric satisfying~\reef{mm}.

The generic 1/8-th supersymmetric $AdS_3$ solution of type IIB supergravity
with non-trivial five-form is then given by taking the metric on $Y_7$
to have the form~\cite{nak1}
\be\label{iibmet}
   ds^2(Y_7) =  \tfrac{1}{4}(dz+P)^2 + e^{-4A} ds^2_6
\ee where $dP={\cal R}$ (the Ricci form on $ds^2_6$). The warp
factor is given by $e^{-4A}=\frac{1}{8}R$ and hence we must have
$R>0$. The five-form flux is given by \bea\label{iibflux1} F_5= L^4
(1+*)\vol(AdS_3)\wedge F \eea with \be\label{iibflux2} F=
\tfrac{1}{2}J-\tfrac{1}{8}d\left[e^{4A}(dz+P)\right] \ee
Using the
fact that the Ricci-form of the K\"ahler metric $ds^2_{2n+2}$
satisfies \be\label{kmlem}
*_{2n+2}{\cal R}=\frac{R}{2}\frac{J^n}{n!}-\frac{J^{n-1}}{(n-1)!}\wedge{\cal R}
\ee
we can rewrite the five-form flux as
\be
F_5=L^4\vol(AdS_3)\wedge F+\frac{L^4}{16}\left[J\wedge {\cal R}\wedge (dz+P)+\frac{1}{2}*_6dR\right]
\ee
since $F$ is clearly closed, we see that $F_5$ is closed
as a result of the condition \reef{mm}.
The vector $\partial_z$ is Killing and preserves the five-form flux.
The solutions are dual to two-dimensional conformal field theories
with $(0,2)$ supersymmetry. Since only the five-form flux is
non-trivial, solutions of this type can be interpreted as arising from
the back-reacted configurations of wrapped or intersecting
D3-branes. For example, we shall show that there are such solutions
that correspond to D3-branes wrapping holomorphic curves in Calabi-Yau
four-folds.

The generic 1/8-th supersymmetric $AdS_2$ solution of $D=11$
supergravity with purely electric four-form flux is given by taking
the internal metric~\cite{nak2}
\bea\label{11met}
ds^2(Y_9) = (dz+P)^2 + e^{-3A}ds^2_8
\eea
with $dP={\cal R}$. The warp factor is $e^{-3A}=\frac{1}{2}R$ and so
again we must take $R>0$.
The four-form flux is given by
\be\label{11flux1}
G_4 = L^3 \vol(AdS_2)\wedge F
\ee
with
\be\label{11flux2}
F=-J+d\left[e^{3A}(dz+P)\right]
\ee
The four-form $G_4$ is clearly closed. Using \reef{kmlem} we have
\be
*G_4=\frac{J^2}{2}\wedge {\cal R}\wedge (dz+P)+\frac{1}{2}*_8dR
\ee
and hence the equation of motion for the four-form, $d*G_4=0$, is satisfied
as a result of \reef{mm}.
Again, the vector $\partial_z$ is Killing and preserves the flux.
These solutions are dual to conformal quantum mechanics with two
supercharges. The fact that the four-form flux is purely electric
means such solutions can be interpreted as arising from the
back-reacted configurations of wrapped or intersecting M2-branes. For
example, as we will see, there are such solutions that
correspond to M2-branes wrapping holomorphic curves in Calabi--Yau
five-folds.

Note for both cases that if we scale the K\"ahler metric by a positive
constant it just leads to a scaling of the overall scale $L$ of the
$D=10$ and $D=11$ solutions.

Finally, we note that a particular class of the $D=11$ solutions can
be related to the type IIB solutions. Suppose that there is a pair of
commuting isometries of the K\"ahler metric $ds^2_8$
such that globally they parametrise a torus $T^2$ and the
nine-dimensional internal manifold is metrically a product
$T^2\times M_7$. By dimensional reduction and T-dualising on this
$T^2$ we can obtain a type IIB solution with an $AdS_2$ factor. In
fact, as we show in the appendix, one actually obtains a type IIB
solution with an $AdS_3$ factor, precisely of the
form~\reef{iibmet}--\reef{iibflux2}.

\subsection{Bubble solutions}

The $AdS$ solutions discussed thus can in general be analytically
continued to describe stationary geometries with
$S^3$ and $S^2$ factors in type IIB and $D=11$ supergravity
respectively. Such ``bubble solutions'' again preserve 1/8-th of the
maximal supersymmetry and generalise the 1/2 supersymmetric bubble
solutions of~\cite{llm}. Generically they have an $\bbR\times SO(4)$
or $\bbR\times SO(3)$ group of isometries. Depending on the boundary
conditions, they can correspond to 1/8-th BPS states in the maximally
SCFTs, or other BPS states in SCFTs with less supersymmetry.

To obtain the type IIB bubble solutions one adapts the analysis
of \cite{nak1}, by replacing the
$AdS_3$ factor with an $S^3$.
The local form of the metric is given by
\be\label{iibmet_bub}
ds^2 = L^2 e^{2A}
\left[-\frac{1}{4}(dt+P)^2 +ds^2(S^3)+ e^{-4A} ds^2_6
\right]
\ee
where $ds^2_6$ is, as before, a K\"ahler metric with Ricci form
$\mathcal{R} = dP$  satisfying~\reef{mm}.
Note that now we have a time-like Killing vector $\partial_t$.
The warp factor is given by $e^{-4A} = -\tfrac{1}{8}R$ and so now we
want solutions with $R<0$. The five-form flux becomes
\be\label{bubf}
F_5 = L^4(1+*)\vol(S^3)\wedge F
\ee
with
\be\label{bubf2}
F = \tfrac{1}{2} J+\tfrac{1}{8} d\left[e^{4A} (dt+P)\right]
\ee
The Killing vector $\partial_t$ also preserves the five-form flux.

Similarly adapting the analysis of~\cite{nak2} by replacing the $AdS_2$
factor by $S^2$ one can construct new bubbling 1/8-th supersymmetric solutions
of $D=11$ supergravity.
The local form of the metric becomes
\bea\label{11metstwo}
ds^2 = L^2e^{2A}\left[-(dt+P)^2 +  ds^2(S^2) +   e^{-3A}ds^2_8\right]
\eea
with $dP={\cal R}$ and the Ricci-form again satisfies~\reef{mm}.
The warp factor is now $e^{-3A}=-\tfrac{1}{2}R$ and so we again want
solutions with $R<0$. The four-form flux is given by
\be\label{11flux1s2}
G_4= L^3 \vol(S^2)\wedge F
\ee
with
\be\label{11flux2s2}
F=-J-d\left[e^{3A}(dt+P)\right]
\ee
and again it is preserved by the Killing vector $\partial_t$.


\section{Fibration constructions using $KE^+_{2n}$ spaces}
\label{fib-con}

In order to find explicit examples of K\"ahler metrics in $2n+2$ dimensions
satisfying~\reef{mm}, we follow~\cite{Page:1985bq} and
also~\cite{Gauntlett:2004yd,Gauntlett:2004hh}, and consider the
ansatz
\be
\label{ansatz}
ds^2_{2n+2} = \frac{d\rho^2}{U} + U \rho^2 (D\phi)^2
   + \rho^2 ds^2(KE^+_{2n})
\ee
with
\be
D\phi=d\phi+B
\ee
Here $ds^2(KE^+_{2n})$ is a $2n$-dimensional Kahler-Einstein metric of
positive curvature. It is normalised so that ${\cal R}_{KE} = 2(n+1)
J_{KE}$ and the one-form form $B$ satisfies $dB=2J_{KE}$. Note
that $(n+1)B$ is then the connection on the canonical bundle of the
K\"ahler-Einstein space.
Let $\Omega_{KE}$ denote a local $(n,0)$-form, unique up rescaling by
a complex function.

To show that $ds^2_{2n+2}$ is a K\"ahler metric observe that the
K\"ahler form, defined by
\be
J = \rho d\rho \wedge D\phi + \rho^2 J_{KE},
\ee
is closed, and that the holomorphic $(n+1,0)$-form
\be
\Omega = e^{i(n+1)\phi}
   \left(\frac{d\rho}{\sqrt{U}}+i\rho\sqrt{U}D\phi\right)
   \wedge \rho^n \Omega_{KE}
\ee
satisfies
\bea
d\Omega  = i f D\phi \wedge \Omega
\label{rp}
\eea
with
\bea
f &=& (n+1)(1-U) - \frac{\rho}{2} \frac{dU}{d\rho}
\eea
This implies, in particular, that the complex structure defined by
$\Omega$ is integrable. In addition~\eqref{rp} allows us to obtain the
Ricci tensor of $ds^2_{2n+2}$:
\be
{\cal R}=dP,\qquad P=fD\phi
\ee
The Ricci-scalar is then obtained via $R={\cal R}_{ij}J^{ij}$.

We would like to find the conditions on $U$ such that $ds^2_{2n+2}$
satisfies the equation~\eqref{mm}. It is convenient to introduce
the new coordinate $x=1/\rho^2$ so that
\begin{equation}
\label{x-ans}
   ds^2_{2n+2} = \frac{1}{x}\left[
      \frac{dx^2}{4x^2U} + U (D\phi)^2 + ds^2(KE^+_{2n}) \right]
\end{equation}
and
\bea
\label{fR}
f &=& (n+1)(1-U)+ x \frac{dU}{dx}\\
R &=& 4n x f - 4x^2 \frac{df}{dx}
\eea
We can now show that~\reef{mm} can be integrated once to give
\be
2n f^2 + U \frac{dR}{dx} = C x^{n-1}
\label{ans}
\ee
where $C$ is a constant of integration.

For simplicity, in what follows, we will only consider polynomial
solutions of \reef{ans}. In particular, if $U(x)$ is a $k$-th order
polynomial we have the following indicial equation:
$(k-n-1)(k-n+1)(2k-n)=0$, which implies that $k=n+1$. Thus our problem
is to find polynomials of the form
\begin{equation}
   U(x) = \sum_{j=0}^{n+1} a_j x^j
\end{equation}
satisfying~\eqref{ans}. Note from~\eqref{fR} that the Ricci scalar is
given by
\begin{equation}
   R = 4x\left[ n(n+1) - \sum_{j=0}^{n-1}(n-j)(n-j+1)a_jx^j \right]
\end{equation}
Since $R$ is related to the $AdS$ warp factor, for a consistent warped
product we see that the range of $x$ must exclude $x=0$. Furthermore,
from \reef{x-ans} we must take $x>0$ and $U(x)>0$.

Our main interest is the six-dimensional ($n=2$) and eight-dimensional
($n=3$) cases, which give rise to type IIB and $D=11$ solutions respectively.
If $n=2$, the function $U(x)$ is cubic and the condition~\reef{ans}
implies that
\bea\label{gop}
a_2^2 - 4a_3^{} a_1^{}&=& 0
\nn
a_3 (1-a_0) &=& 0
\nn
(a_0-1)(a_0-3) &=& 0
\eea
The Ricci scalar is given by
\begin{equation}
\label{n3-R}
   R = 8x \left( 3 - 3a_0 - a_1x\right)
\end{equation}
We have two choices depending on the solution of the second
equation of \reef{gop}. The first is $a_3=0$ and hence $a_2=0$ with
\be
U(x)=a_0+a_1x
\ee
and either $a_0=1$ or $a_0=3$.
The second choice is $a_3\neq0$, $a_0=1$ and hence
$a_2^2=4a_3 a_1$ with
\be\label{cubu}
U(x)=1+a_1x+a_2x^2+(a_2^2/4a_1)x^3
\ee
(note that when $a_1=a_2=0$ we have $R=0$ and so we ignore this case.)

If $n=3$ the function $U(x)$
is quartic and the condition~\reef{ans} implies that
\bea\label{gop2}
a_3^2 - 4a_4^{} a_2^{} &=& 0 \nn
a_4 a_1 &=& 0 \nn
a_1(2-a_0) &=& 0 \nn
a_3a_1 - 4a_4(1-a_0)&=&0 \nn
(a_0-1)(a_0-2) &=& 0
\eea
The Ricci scalar is given by
\begin{equation}
\label{3nRicci}
   R = 8x\left( 6 - 6a_0 - 3a_1x - a_2x^2\right)
\end{equation}
Solving the equations \reef{gop2} again leads to two classes of solutions
depending on the solution of the second equation. First we take
$a_4=0$ which implies $a_3=0$ with
\be
U(x) = a_0+a_1x+ a_2x^2
\ee
and either $a_0=1$, $a_1=0$ or $a_0=2$.
Alternatively we take $a_4\neq0$, $a_1=0$
and hence $a_0=1$ and $a_3^2=4a_4a_2$ with
\be
U(x) = 1+a_2x^2+a_3x^3+(a_3^2/4a_2)x^4
\ee
(note that when $a_0=1,a_1=a_2=0$ we have $R=0$ and so we ignore this case.)

For the remainder of this section we will only consider $AdS$ solutions ($R>0$), returning to bubble
solutions ($R<0$) in the next section. In order that the $AdS$ solutions
are globally defined, in the following we will
usually assume that the local leaves in~\eqref{x-ans} with metric
$ds^2(KE^+_{2n})$ extend globally to form a compact K\"ahler--Einstein
manifold $KE^+_{2n}$ and that the internal manifold $Y_{2n+3}$ in \reef{firsteq}
is a
fibration over $KE^+_{2n}$. We could also assume that $x$ and $\phi$
in~\eqref{x-ans} separately describe a fibration over $KE^+_{2n}$. In
particular, if the range of $x$ is taken to lie between two zeroes
of $U(x)$, then, at a fixed point on $KE^+_{2n}$, $(x,\phi)$ can
parametrise a two-sphere ($U(x)$ has to have a suitable behaviour at the
zeroes to avoid conical singularities). Topologically this can then
form a two-sphere bundle over $KE^+_{2n}$ which is just the canonical
line bundle $KE^+_{2n}$ with a point ``at infinity'' added to each of
the fibres. In fact we shall see that in the solutions we discuss this
possibility is {\it not} realised and that the metric necessarily has
conical singularities at one of the poles of the two-sphere. However,
as we shall also see, in the $D=10$ and $D=11$ supergravity solutions, after adding in the
extra $z$-direction, in the resulting spaces $Y_{2n+3}$
two-sphere bundles over $KE^+_{2n}$ (without conical
singularities) do appear but where the polar angle on the sphere is a
combination of $\phi$ and $z$.

We will now discuss the six-dimensional ($n=2$) and eight-dimensional
($n=3$) cases in turn, corresponding to type IIB $AdS_3$ and $D=11$
$AdS_2$ solutions respectively.

\subsection{Fibrations over $KE_4^+$: type IIB $AdS_3$ solutions}
\label{fib-iib}

For these solutions, the warp factor is given by
$R=8e^{-4A}$ which must be positive.
Recall that we had two choices for the function $U(x)$.
First consider the case $U(x)=a_0+a_1x$ with either $a_0=1$ or $a_0=3$.
From~\eqref{x-ans} for a compact $Y_7$
manifold with finite warp factor we need a finite range of $x>0$ between
two solutions of $U(x)=0$, such that $U(x)>0$ (so that we
have the right signature). Since $U(x)$ is linear for this case, it has only one root
and so there are no compact solutions. (In fact, as we discuss later,
the case where $a_0=1$ corresponds to $AdS_5\times
X_5$, where $X_5$ is a Sasaki--Einstein manifold.)

We thus consider the second case $U(x)=1+a_1x+a_2x^2+(a_2^2/4a_1)x^3$.
We now show that this gives rise to the family of
type IIB $AdS_3$ solutions found in~\cite{gmmw2}.
To compare with the solutions given in~\cite{gmmw2} we need to make a
number of transformations. First it is convenient to change
parametrization and write $a_3=-1/\alpha^3,a_2=2\beta/\alpha^3,
a_1=-\beta^2/\alpha^3$
so that
\be
U(x) = 1 - \frac{x(x-\beta)^2}{\alpha^3}.
\label{eq_u}
\ee
The scalar curvature is given by
\be
R = \frac{8\beta^2 }{\alpha^3} x^2
\ee
and we must choose $\alpha>0$ to ensure that $R>0$.
The metric~\eqref{iibmet} on the internal manifold is then given by
\be\label{notcon}
ds^2(Y_7) =
   \frac{1}{4}\left[dz- \frac{2\beta x(x-\beta)}{\alpha^3} D\phi\right]^2
   + \frac{\beta^2}{\alpha^3} \left[
      \frac{dx^2}{4xU} + xU (D\phi)^2 + x ds^2(KE_4^+)
   \right]
\ee
Note that this is invariant under simultaneous rescalings of $x$,
$\alpha$ and $\beta$. Using this symmetry we can set
\be
\beta=\frac{4}{3a},\qquad \alpha^3=\frac{256}{729 a^2}
\ee
Introducing new coordinates $y=4/(9x)$ and $\psi=3\phi+z$ we can
rewrite the metric as
\begin{equation}
\label{soloth}
   ds^2(Y_7) =\frac{y^2-2y+a}{4y^2}Dz^2+ \frac{9dy^2}{4q(y)} +
      \frac{q(y) D\psi^2}{16y^2(y^2-2y+a)} +
      \frac{9}{4y} ds^2(KE_4^+)
\end{equation}
where $D\psi=d\psi+3B$, $Dz=dz-g(y)D\psi$ and
\begin{equation}
\begin{aligned}
   q(y) &= 4y^3-9y^2+6ay-a^2 \\
   g(y) &= \frac{a-y}{2(y^2-2y+a)}
\end{aligned}
\end{equation}
The warp factor is simply $e^{2A}=y$. Using~\reef{iibflux1}
and~\reef{iibflux2}, we find the five-form flux is
\be
F=-\frac{1}{4}ydy\wedge dz +\frac{3a}{8} J_{KE}
\ee
This can now be directly compared with the solution constructed
in~\cite{gmmw2}. We need to take into account the different
normalization conventions for the $KE_4^+$. This requires rescaling
$ds^2(KE^+_4)$ by a factor of $1/6$. Recalling that by definition
$3B$ is the connection on the canonical bundle of $KE_4^+$,
we see that~\eqref{soloth} agrees precisely with the metric
given in~\cite{gmmw2}. Furthermore, the expression for the five-form
also agrees, again allowing for a difference in conventions: the five
form being used here is $-1/4$ that of \cite{gmmw2}.

The regularity of these solutions were discussed in detail
in~\cite{gmmw2}. Restricting $y$ to lie between the two smallest roots
of the cubic $q(y)$, topologically the solutions as written
in~\reef{soloth} are $U(1)$ bundles, with fibre parametrised by $z$,
over an $S^2$ bundle, with fibre parametrised by $(\psi,y)$, over
$KE_4^+$. Note also, as was mentioned above, that the six-dimensional
K\"ahler leaves parametrised by $(x,\phi)$ and $KE_4^+$ appearing
in~\reef{notcon} are not $S^2$ bundles over $KE_4^+$ as there is
necessarily a conical singularity at one of the poles.

\subsection{Fibrations over $KE_6^+$: $D=11$, $AdS_2$ solutions}

We now discuss the case where $n=3$ and $Y_9$ is nine-dimensional and
look for $AdS_2$ solutions to $D=11$ supergravity.
For these solution the warp factor is given by $R = 2e^{-3A}$.
Recall that there were two choices for the function $U(x)$.
First consider $U(x) = a_0+a_1x+a_2x^2$.
Recall again that for a compact $Y_9$ geometry with finite warp factor
we need to find a range of $x>0$ between the two roots of $U(x)=0$ over
which $U(x)>0$. Since $a_0=1$ or $a_0=2$, this is not possible and
thus there
are no compact warped product solutions in this class. (In fact, as
we discuss later, the case $a_0=1$ corresponds to $AdS_4\times X_7$
where $X_7$ is Sasaki--Einstein.)

We thus focus on the second case for which
$U(x) = 1+a_2x^2+a_3x^3+(a_3^2/4a_2)x^4$.
This implies $R=-8a_2x^3$.
Since we require $x>0$, $R>0$ we must have $a_2<0$ and hence
$a_4<0$. It is then useful to redefine $a_4= -1/\alpha^4$,
$a_3=4\beta/\alpha^4$ and $a_2=-4\beta^2/\alpha^4$, with $\alpha>0$, so that
\be\label{yoohoo}
U(x) = 1- x^2(x-2\beta)^2/\alpha^4
\ee
and
\be
R = \frac{32\beta^2}{\alpha^4} x^3
\ee
The metric on the internal space $Y_9$ is then given by
\be\label{dipsy}
ds^2(Y_9) = (dz+P)^2 + \frac{16\beta^2}{\alpha^4}
\left[ \frac{dx^2}{4U} + x^2U (D\phi)^2 + x^2 ds^2(KE_6^+)
\right]
\ee
where $P=-4\alpha^{-4}\beta x^2(x-2\beta)D\phi$. Note that the metric
is invariant under simultaneous rescalings of $x$, $\beta$ and
$\alpha$.
The roots of $U(x)=0$ are given by
\begin{equation}
\begin{aligned}
   x_1 &= \beta - \sqrt{\beta^2+\alpha^2} &\qquad
   x_2 &= \beta - \sqrt{\beta^2-\alpha^2} \\
   x_3 &= \beta + \sqrt{\beta^2-\alpha^2} &\qquad
   x_4 &= \beta + \sqrt{\beta^2+\alpha^2}
\end{aligned}
\end{equation}
Note that for $\beta^2>\alpha^2$ we have four real roots and $U(x)\geq0$
for $x\in[x_1,x_2]$ and $x\in[x_3,x_4]$. Demanding that $x>0$, $R>0$ we deduce that
$\beta^2>\alpha^2$, $\beta>0$ and $x\in[x_3,x_4]$.

As in the type IIB case, it is not possible to avoid
conical singularities at both $x_3$ and $x_4$ just by adjusting the
period of $\phi$. However, we can take an appropriate linear
combination of $z$ and $\phi$ and find a smooth compact manifold.
To this end, we change coordinates
\be
\psi = 4\phi + \zG
\ee
It is also convenient to use the scaling symmetry to set
\be
\beta=\frac{3^{3/2}}{2^{11/2}a},\qquad \alpha^4=\frac{3^6}{2^{22}a^2}
\ee
and change variables from $x$ to $y=3^{3/2}/(2^{11/2}x)$. In these
coordinates the warp factor is simply
\be
e^{2A}=\frac{2}{3}y^2 .
\ee
With these changes the metric takes the form
\begin{equation}
\label{q-met}
\begin{aligned}
   ds^2(Y_9) = \frac{y^3-3y+2a}{y^3}Dz^2
     + \frac{4dy^2}{q(y)}+\frac{q(y)(D\psi)^2}{y^3(y^3-3y+2a)}
      + \frac{16}{y^2}ds^2(KE_6^+)
\end{aligned}
\end{equation}
where $D\psi=d\psi+4B$, $Dz=d\zG-g(y)D\psi$ and
\begin{equation}
\begin{aligned}
   q(y) &= y^4 - 4 y^2 + 4 a y - a^2 \\
   g(y) &= \frac{a-y}{y^3 - 3 y + 2a} .
\end{aligned}
\end{equation}
The conditions $\beta>0$ and $\beta^2>\alpha^2$ translate into
$0<a<1$. The function $U(x)$ has been replaced by $q(y)$, which again
has four roots $y_1<0<y_2<y_3<y_4$, for this range of $a$. The
condition that $x\in[x_3,x_4]$ translates into $y\in[y_2,y_3]$.

Near a root $y=y_i$ we find that the $(y,\psi)$ part of the metric is
given by
\begin{equation}
   \frac{16}{q'(y_i)}\left[
      dr^2 + \frac{q'(y_i)^2}{16y^3(y^3-3y-2a)} (D\psi)^2 \right]
   = \frac{16}{q'(y_i)}\left[
      dr^2 + r^2 (D\psi)^2 \right]
\end{equation}
where $y-y_i=r^2$. Thus, remarkably, by choosing the period of $\psi$
to be $2\pi$ we can avoid conical singularities at both $y=y_2$ and
$y=y_3$. As a consequence we can look for solutions where the topology
of $Y_9$ is a $U(1)$ bundle, whose fibre is parametrised by $\zG$,
over an eight-dimensional manifold which is topologically a two-sphere
bundle, parametrised by $(y,\psi)$, over $KE_6^+$. Furthermore, since
by definition $4B$ is the connection of the canonical bundle of $KE_6^+$,
the two-sphere bundle is simply the canonical
line bundle of $KE_6^+$ with a ``point at infinity'' added to each
fibre.

In order to check that the $U(1)$ fibration, with fibre parametrised
by $\zG$, is globally defined, we need ensure that the periods of
$d(gD\psi)$ over all 2-cycles of the eight-dimensional base space are
integer valued.  The problem is very similar to the type IIB solutions
and we can follow the analysis of~\cite{gmmw2}. If we let the period
of $\zG$ be $2\pi l$, then we must have
\be
g(y_3)-g(y_2)=lq,\qquad g(y_2)=lp/m
\ee
for some integers $p$ and $q$. The integer $m$ is fixed by the
choice of $KE_6^+$ manifold: if $\mathcal{L}$ is the canonical line
bundle, then $m$ is the largest possible positive integer such that
there exists a line bundle ${\cal N}$ with
$\mathcal{L}=\mathcal{N}^m$. Furthermore, if the integers $p$ and $q$
are relatively prime $Y_9$ is simply connected. These conditions imply
that we must choose
\be
a=\frac{mq(2p+mq)}{(2p^2+2mpq+m^2q^2)}
\ee
and
\be
l^2=\frac{m^2(2p^2+2mpq+m^2q^2)}{2p^2(p+mq)^2}
\ee

Finally, we note that the four-form flux is given by \reef{11flux1} with
\be
F=\frac{2^{3/2}}{3^{3/2}}\left[3y^2dy\wedge d\zG-8aJ_{KE}\right]
\ee
It is straightforward to determine the additional conditions
imposed by demanding that the four-form is properly quantised but
we shall not do that here.

\subsection{Non-compact $AdS_2$ solutions in $D=11$ and defect CFTs}

Given the solutions~\eqref{q-met}, we can return to the original
angular variables $\phi$ and $z$ and complete the squares in a
different way, so the eleven-dimensional metric reads
\begin{equation}
\begin{aligned}
   ds^2 &= \frac{2y^2}{3}ds^2(AdS_2) + \frac{32}{3}ds^2(KE_6^+)
      \\ &\qquad\qquad
      + \frac{32}{3}\left[D\phi
        + \left(\frac{1}{2}-\frac{a}{4y}\right)dz\right]^2
      + \frac{8y^2}{3q(y)}dy^2+\frac{2q(y)}{3y^2}dz^2
\end{aligned}
\end{equation}
Let us now consider letting the range of $y$ be given by $y_4\le y\le
\infty$, where $y_4$ is the largest root of the quartic
$q(y)$. Clearly this gives rise to non-compact solutions with $AdS_2$
factors. These are the analogue of the non-compact $AdS_3$ solutions
of type IIB supergravity that were discussed in section 7 of
\cite{gmmw3}.

Observe that when $a=0$, after implementing the coordinate change
$y^2=4\cosh^2 r$ and $\phi'=4\phi+2z$ we obtain
\be
\tfrac{3}{8}ds^2 = \cosh^2 r ds^2(AdS_2) + dr^2 + \sinh^2 r dz^2
   +4\left[ ds^2(KE_6^+) + \tfrac{1}{16}(d\phi'+ 4B)^2 \right]
\ee
This is simply the $AdS_4\times SE_7$ solution of $D=11$ supergravity
where $SE_7$ is a seven-dimensional Sasaki-Einstein manifold. In
particular, in the special case that we choose $KE_6^+$ to be $\bbC P^3$,
we get the standard $AdS_4\times S^7$ solution. Note that
if $SE_7$ is regular or quasi-regular, then $KE_6^+$ is a globally defined
manifold or orbifold, respectively, while if it is irregular,
$KE_6^+$ is only locally defined.

We next observe that for general $a$, as $y\to\infty$ the solution
behaves as if $a=0$ and hence the solutions are all asymptotic to
$AdS_4\times SE_7$. By choosing the period of the coordinate $z$
suitably, we can eliminate the potential conical singularity as $y$
approaches $y_4$. With this period the non compact solutions are
regular: they are fibrations of $SE_7$ over a four-dimensional space
which is a warped product of $AdS_2$ with a disc parametrised by
$(y,z)$.

To interpret these solutions we consider for simplicity the case when
$SE_7=S^7$. Now, there are probe membranes in $AdS_4\times S^7$ whose
world-volume is $AdS_2\times S^1$. More precisely, the $AdS_2$
world-volume is located in $AdS_4$ while the $S^1$ is the Hopf fibre
of $S^7$. These configurations preserve 1/16-th of the Minkowski
supersymmetry and are a generalisation\footnote{In~\cite{st} probe
  membranes with world-volume $AdS_2\times S^1$ were also considered
  but they are not the same as those being considered here as they
  preserve 1/4 of the supersymmetry.} of those studied in~\cite{kr}
corresponding to defect CFTs. It is natural therefore to interpret our
new solutions as the back reacted geometry of such probe branes. One
might expect that the back reacted geometry of such branes to be
localised in $\bbC P^3$, however, in our solutions the $\bbC P^3$ is
still manifest. Hence our geometries seem to correspond to such probe
membranes that have been ``smeared'' over the $\bbC P^3$.

We make a final observation about the $a=1$ case, for which $q(y)$ has
a double root at $y=1$. By expanding the solution near $y=1$ we find
that the solution is asymptotically approaching the solutions
discussed in section~\reef{gensuk} below. In particular, for the
special case when $KE_6^+=\bbC P^3$, this is the solution found
in~\cite{gkwp} that describes the near horizon limit of membranes
wrapping a holomorphic $H^2/\Gamma$ in a Calabi--Yau five-fold. Thus,
in this special case, our full non-compact solution, interpolates
between $AdS_4\times S^7$ and the solution of~\cite{gkwp}, while
preserving an $AdS_2$ factor. Note that this is entirely analogous to
the discussion of the non-compact type IIB $AdS_3$ solutions discussed
in section 7 of~\cite{gmmw3}.


\section{Bubbles from fibrations over $KE_{2n}^+$ and Superstars}

We will now use the same local K\"ahler metrics described at the
beginning of section~\ref{fib-con} to construct supersymmetric bubble
solutions with $S^3$ factors in type IIB and $S^2$ factors in
$D=11$. The key point is simply to consider a different range of the
variable $x$ such that Ricci scalar $R$ is now negative.

\subsection{Type IIB solutions from fibrations over $KE^+_4$}

We first observe that if we take $U(x)=1+a_1x$, with $a_1>0$ to ensure
that $R<0$, and choose the four-dimensional K\"ahler--Einstein base,
$KE^+_4$, to be $\CP^2$ we simply recover the $AdS_5\times S^5$
solution. This becomes clear after making the coordinate
transformation $\phi\to \phi-\tfrac{1}{2}t$. More generally by taking
the same $U$ but with different choices of $ds^2(KE^+_4)$ metric
(note the corresponding leaves need not extend globally to form a
compact K\"ahler--Einstein space)  we can  obtain an $AdS_5\times
SE_5$ solution, where $SE_5$ is any arbitrary five-dimensional
Sasaki--Einstein manifold.

Let us now consider the solutions based on the more general
cubic~\reef{cubu}.
Since taking $a_2=0$ returns to the $AdS_5\times SE_5$ described
above, we expect these solutions to correspond to excitations in the
CFT dual of the Sasaki--Einstein solutions. We again must have $a_1>0$
to ensure $R<0$. It is convenient to rescale the coordinate $x$ so
that $a_1=1$ (this leads to an overall scaling of the six-dimensional
K\"ahler metric which can be absorbed into the overall scale $L$ of
the $D=10$ metric) giving
\be
U=1+x+a_2x^2+\tfrac{1}{4}a_2^2x^3.
\ee
If make the change of coordinate $x=1/(r^2+Q)$ where $Q=-a_2/2$ we find
that the type IIB metric can be written as
\be
ds^2=-\tfrac14H^{-2}f\,d t^2 + H
   \left[ f^{-1}d r^2+r^2d s^2(S^3)\right]
   + d s^2(KE^+_4) +  (D\phi+A)^2
\ee
where
\bea
H&=&1+\frac{Q}{r^2}\nn
f&=&1+r^2H^3\nn
A&=&\tfrac12 H^{-1}dt .
\eea

When $KE^+_4=\CP^2$ we see that this is precisely the single-charged $AdS_5$
``black hole'' solution given in~\cite{London:1995ib,Behrndt:1998ns}, uplifted to
$D=10$ using an $S^5$, as described in~\cite{Cvetic:1999xp}. The fact
that we can replace the $\CP^2$ with any $KE^+_4$ is a consequence of
the recent result that there is a consistent Kaluza-Klein truncation
to minimal $D=5$ gauge supergravity using any $D=5$ Sasaki-Einstein
space~\cite{Buchel:2006gb}. These $D=10$ solutions were interpreted as
corresponding to giant gravitons and were called superstars
in~\cite{Myers:2001aq}.

\subsection{$D=11$ solutions from fibrations over $KE^+_6$}

We now start with $U=1+a_2x^2$ with $a_2>0$. If we choose
$ds^2(KE^+_6)$ to be the metric on $\CP^3$ it is again easy to show
that one recovers the $AdS_4\times S^7$ solution. More generally we
get $AdS_4\times SE_7$ solutions for arbitrary Sasaki--Einstein
seven-manifold $SE_7$ for suitable different choices of the local
metric $ds^2(KE^+_6)$.

As before, solutions based on the more general quartic~\reef{yoohoo}
should then correspond to excitations in the CFT dual of the
Sasaki--Einstein solutions. Scaling $x$ so that $a_2=1$, we have
\be
U=1+x^2+a_3x^3+\frac{a_3^2}{4}x^4
\ee
If we make the change of variable $x=1/(r+Q)$ where $Q=-a_3/2$ we find that
the $D=11$ metric can be written as
\be
4^{2/3}ds^2=
-H^{-2}fdt^2+H^2\left[f^{-1}dr^2+r^2 ds^2(S^2)\right]
+4ds^2(KE^+_6)+4\left(D\phi+\tfrac{1}{2}A\right)^2
\ee
where
\bea
H&=&1+\frac{Q}{r}\nn
f&=&1+r^2H^4\nn
A&=&H^{-1}dt
\eea

When $KE^+_4=\CP^3$ this is precisely the supersymmetric single-charged
$AdS_4$ ``black hole'' discussed in~\cite{Romans:1991nq}, uplifted to
$D=11$ using an $S^7$ as described in~\cite{Cvetic:1999xp}. The fact
that we can replace the $\CP^3$ with any $KE^+_6$ is very suggestive
that there is a consistent Kaluza-Klein truncation to minimal $D=4$
gauge supergravity using any $D=7$ Sasaki-Einstein space (thus
generalising the result of \cite{Buchel:2006gb}; see
also~\cite{Gauntlett:2006ai}). These $D=11$ solutions were interpreted
as corresponding to giant gravitons and were called superstars in~\cite{Leblond:2001gn}.

\section{New $AdS$ solutions from multi-charged superstars}

In this section we will derive new $AdS$ solutions from the
general three-charged and four-charged superstar solutions of type IIB and $D=11$,
respectively. The strategy is to first identify the K\"ahler geometry
underlying the superstar solutions and then by a judicious rescaling
and change of variables, construct a K\"ahler metric with
positive Ricci scalar and use this to build the $AdS$ solutions.

\subsection{Type IIB three-charged superstars}

We start by summarising the three-charge superstar geometry as
presented in~\cite{Cvetic:1999xp}. If we relate our time coordinate
$t$ to the time coordinate $\tilde{t}$ of that reference by
$\tilde{t}=\tfrac12t$, we find that the solution can be put in the
bubble form~\reef{iibmet_bub} with
\bea
\label{Cv-bub}
e^{4A}&=&{\cal D}{\cal H}r^4\nn
P &=&\frac{2}{r^2{\cal DH}}\sum_i \mu^2_i d\phi_i \nn
ds^2_6&=&\frac{{\cal DH}r^2}{f}dr^2+
r^2\sum_i H_i(d\mu^2_i+\mu_i^2d\phi^2)
+\frac{1}{{\cal DH}}\Big(\sum_i \mu^2_i d\phi_i\Big)^2
\eea
where with $i=1,2,3$
\begin{equation}
   H_i = 1 + \frac{Q_i}{r^2}
\end{equation}
and we have defined
\bea
{\cal H} &=& H_1 H_2 H_3 \nn
f&=&1+r^2{\cal H}\nn
{\cal D} &=& \sum_i \frac{\mu^2_i}{H_i}
\eea
Furthermore if we write the five-form, $\tilde{F}_5$
of~\cite{Cvetic:1999xp} as $\tilde{F}_5=4F_5$,
then we find\footnote{Note that there is a typo in the sign of the
second term in equation (2.8) in \cite{Cvetic:1999xp}.}
 that it takes the form~\reef{bubf} and~\reef{bubf2}
with
\be
F=\tfrac{1}{8}d\left[e^{4A}(d t+P)\right]+\tfrac12J
\ee
where
\be
J=-rdr\wedge\sum_i\mu^2_i d\phi_i
-\frac{r^2}{2}\sum_i H_id(\mu^2_i)\wedge d\phi_i
\label{kahler6}
\ee
It is easy to see that this form is closed. To see that
it is indeed a K\"ahler form corresponding to the metric $ds^2_6$
it is convenient to choose the orthonormal frame as
\begin{equation}
\label{basis}
\begin{aligned}
   e^i &= \frac{r {\cal H}^{1/2}}{f^{1/2}} \frac{\mu_i}{H^{1/2}_i} dr
      + r H^{1/2}_i d\mu_i \\
   \tilde{e}^i &= \frac{C}{{\cal D}} \frac{\mu_i}{H^{1/2}_i}
      \sum_j \mu^2_j d\phi_j + r H^{1/2}_i \mu_i d\phi_i
\end{aligned}
\end{equation}
One can then show that the metric can be written as
\be
ds^2_6 = \sum_i \left(e^i \otimes e^i + \tilde{e}^i \otimes \tilde{e}^i \right)
\ee
provided that $C$ satisfies
\be
(C+r)^2 = r^2+ {\cal H}^{-1} = \frac{f}{{\cal H}}
\ee
In this frame, we find that $J$ takes the conical form:
\be
J = -\sum_i e^i \wedge \tilde{e}^i
\ee
It is possible to directly show that the relevant complex
structure is integrable, which completes a direct confirmation that
the metric $ds^2_6$ is indeed K\"ahler (of course this is all
guaranteed since the solutions of~\cite{Cvetic:1999xp} are known to
preserve 1/8 supersymmetry).
As a further check one can calculate the Ricci-form
from the expression for $P$ and, using the expression for $J$, the
Ricci scalar. One can then compare with the expression for the warp
factor and check that we have $R=-8e^{-4A}$.

\subsection{Type IIB $AdS_3$ solutions}
\label{SSadS}

We now want to use the six-dimensional K\"ahler metrics coming from
the three-charge superstars to construct new $AdS_3$ solutions. As
stands it is not immediately obvious how to do so since these metrics
have Ricci curvature $R<0$ whereas for $AdS_3$ solutions we need
$R>0$. Recall that in the equal charge case the two types of solution
arose from the same more general class of K\"ahler metrics with
$U(x)=1+a_1x+a_2x^2+(a_2^2/a_1)x^3$. These had $R=-8a_1x^2$. Rescaling
the coordinate $x$ we could set either $a_1=-1$ which led to $AdS_3$
solutions or $a_1=1$ leading to superstar solutions. Clearly if we
want to use the three-charge superstar solutions to construct new
$AdS_3$ geometries we need to extend the solutions by introducing the analogue
of the $a_1$ parameter.

This is easy to do simply by using the scale invariance of solutions. The
condition on the curvature~\eqref{mm} is clearly invariant under
constant rescalings of the metric $ds_{2n+2}^2$ (as are of course the
K\"ahler conditions). We know that the metric~\eqref{Cv-bub} is
K\"ahler and satisfies~\eqref{mm} with Ricci scalar
$R=-\mathcal{DH}r^4$. Thus, from the rescaling symmetry, $\lambda
ds^2_6$ is also a solution. Now consider making the change of
variables $w=\lambda r^2$ and defining new parameters $q_i=\lambda
Q_i$, so $H_i=1+q_i/w$. The rescaled metric can then be written as
\begin{equation}
\label{lmetric}
   ds^2_6 = \frac{Y}{4F}dw^2+\sum_i(w+q_i)(d\mu^2_i+\mu^2_id\phi_i^2)
          +\frac{F-1}{Y}\Big(\sum_i \mu^2_i d\phi_i\Big)^2
\end{equation}
where we have introduced
\begin{equation}
   \begin{aligned}
      Y(w) &= \sum_i\mu^2_i(w+q_i)^{-1} \\
      F(w) &= 1+\lambda w^2\prod_i (w+q_i)^{-1}
   \end{aligned}
\end{equation}
while the scalar curvature and $P$ are given by
\begin{equation}
\begin{aligned}
   R &= - \frac{8(F-1)}{w^2Y} \\
   P &= \frac{2(F-1)}{wY}\sum_i\mu^2_i d\phi_i
\end{aligned}
\end{equation}
One notes the similarity in the parametrization with the form of the
seven-dimensional Sasaki--Einstein metrics given in~\cite{Cvetic:2005ft}.
Note also that the new scale factor parameter $\lambda$ appears only in
$F(w)$.

The original superstar solution corresponded to $\lambda=1$ with
$R<0$. For the $AdS_3$ we instead take $\lambda=-1$ with $R>0$. For
the metric to be positive definite we need to choose the range of $w$
so that $w+q_i>0$. This implies that $Y>0$ and $F<1$ so that $R>0$ as
required. Finally for the first term in~\eqref{lmetric} to be
well-defined we also require $F\ge 0$. This can be achieved by
choosing suitable values of $q_1\le q_2\le q_3$ so that the cubic $\prod
(w+q_i)-w^2$ has three zeroes $w_1<w_2<w_3$ and demanding that $w_1\le
w\le w_2$ with $w_1>-q_1$. Given that $F-1<0$ it is not obvious that
the metric is in fact positive definite. In the equal charge case, it is
straightforward to show that it in fact is. Rather than show it for
the general case, let us instead examine $ds^2(Y_7)$:
\bea
\label{Y7}
ds^2(Y_7)&=&\tfrac{1}{4}(dz+P)^2+e^{-4A}ds^2_6 \\
&=&\frac{F}{4}dz^2
+\frac{1-F}{4w^2F}dw^2+\frac{1-F}{Yw^2}\sum_i\left(w+q_i\right)
\bigg[d\mu^2_i+\mu^2_i\Big(d\phi_i-\frac{wdz}{2(w+q_i)}\Big)^2
\bigg] \nonumber
\eea
which is clearly positive definite.

To analyse the global structure of these metrics, we follow the approach
of~\cite{Cvetic:2005ft}. We first observe that the metrics are
co-homogeneity three with $U(1)^4$ principle orbits which will
degenerate at various points. The four local isometries are generated
by $\del_z$ and $\del_{\phi_i}$. Globally we would like to find
combinations of these Killing vectors which generate compact
$U(1)$ orbits.

From the form~\eqref{Y7} we see there are degenerations
at $\mu_i=0$ and also at $F=0$. For the former, the Killing vector
whose length is vanishing is simply $\del_{\phi_i}$. It is easy to see
that for the metric to be smooth at $\mu_i=0$ we require $\phi_i$ to
have period $2\pi$. For the degenerations at roots $w=w_1$ and $w=w_2$
of $F$ the Killing vector whose length is vanishing is given by
\begin{equation}
   l_i = c_i \del_z + c_i \sum_j \frac{w_i}{2w_i+2q_j}\del_{\phi_j} ,
\end{equation}
for $i=1,2$ and some constant $c_i$. The requirement of regularity of
the metric at these points can found either by requiring that $l_i$ is
normalised so that corresponding surface gravity $\kappa_i$ is unity
\be
\kappa_i^2=\frac{g^{\mu\nu}\partial_\mu (l_i^2) \partial_\nu (l_i^2)}
{4l_i^2}
\ee
or by direct inspection of the metric by introducing a coordinate
corresponding to $l_i$. In this case the latter is
relatively straightforward and one finds that the constants $c_i$ must
be given by
\begin{equation}
   c_i^{-1} = -1+\sum_j \frac{w_i}{2w_i+2q_j}
\end{equation}
which is again very similar to the conditions
in~\cite{Cvetic:2005ft}.

We now have found conditions arising from five different
degenerations: the three points $\mu_i=0$ together with $w=w_1$ and
$w=w_2$. However, there are only four isometries so there must be a
relation of the form
\begin{equation}
   pl_1+ql_2+\sum_j r_j\partial_{\phi_j}=0
\end{equation}
for some co-prime integers $(p,q,r_i)$. This then further restricts
the parameters $q_i$. Since we can have $U(1)^2$ degenerations when
$w=w_i$ and $\mu_j=0$, we also require $p$ and $q$ are separately
coprime to each of the $r_i$.

To ensure that we have a good solution of type IIB string theory we should
also ensure that the five-form is suitably quantised. We will leave
a detailed analysis of this issue for future work.

\subsection{$D=11$ four-charged superstars}

Turning to solutions of $D=11$ supergravity, we start by summarising
the four-charged superstar geometry as an example of a 1/8 BPS state
with an $S^2$ factor.  In the next subsection we then adapt the
metric as in the previous discussion to give a new class of $AdS_2$
solutions.

We first put the solution into our standard bubble form~\eqref{11metstwo}
starting from the form presented in~\cite{Cvetic:1999xp}. We
find, setting $g=\frac12$ in~\cite{Cvetic:1999xp}, that
\begin{equation}
\begin{aligned}
e^{3A}&=
   \mathcal{DH}r^3 \\
P&=
   \frac{2}{r^2\mathcal{DH}}\sum_i \mu^2_i d\phi_i \\
ds^2_8 &=
   \frac{\mathcal{DH}r}{f}d r^2
   + 4r \sum_i H_i \left( d\mu_i^2 + \mu_i^2 d\phi_i^2 \right)
   + \frac{4}{\mathcal{DH}r}\Big( \sum_i \mu^2_i d\phi_i \Big)^2
\end{aligned}
\end{equation}
where for $i=1,\dots,4$
\begin{equation}
   H_i = 1 + \frac{Q_i}{r}
\end{equation}
and we have defined
\bea
{\cal H} &=& H_1 H_2 H_3 H_4 \nn
f&=&1+r^2{\cal H}\nn
{\cal D} &=& \sum_i \frac{\mu^2_i}{H_i}
\eea
Furthermore the four-form flux\footnote{Note that there is a typo in the sign of the
third term in equation (3.6) of \cite{Cvetic:1999xp}.} takes the form~\eqref{11flux2s2} with
\be
F = d\left[e^{3A}(d t+P)\right] + J
\ee
where
\be
J = -2dr\wedge\sum\mu^2_i d\phi_i-2r\sum H_id(\mu^2_i)\wedge d\phi_i
\ee
which is clearly closed. To show that the metric $ds_8^2$ is indeed
K\"ahler  with K\"ahler form $J$, one can introduce an orthonormal
frame in analogy with~\eqref{basis}. Again, one could also check that
the corresponding curvature satisfies~\eqref{mm} and that $R=-2e^{-3A}$.

\subsection{$D=11$ $AdS_2$ solutions}

As before we can use the rescaling symmetry of the K\"ahler metric
defining the four-charged superstar metric to obtain new $D=11$
$AdS_2$ solutions. The internal space for the $AdS_2$
solutions we have already considered are analogous to the
nine-dimensional Sasaki-Einstein spaces constructed
in~\cite{Gauntlett:2004hh}. The new $AdS_2$ solutions we now consider
are then analogous to the nine-dimensional Sasaki-Einstein spaces
considered in~\cite{Cvetic:2005ft}.

We define $w=\lambda r$, rescale the metric by $\lambda$, define
$q_i=\lambda Q_i$ and introduce as before
\begin{equation}
\begin{aligned}
   Y(w) &= \sum_i\mu^2_i(w+q_i)^{-1} \\
   F(w) &= 1 + \lambda^2 w^2\prod (w+q_i)^{-1}
\end{aligned}
\end{equation}
We then find that the rescaled metric is given by
\begin{equation}
ds^2_8 =  \frac{Y}{F}dw^2+4\sum_i(w+q_i)(d\mu^2_i+\mu^2_id\phi^2)
 +\frac{4(F-1)}{Y}\bigg(\sum_i\mu_i^2d\phi_i\bigg)^2
\end{equation}
with
\begin{equation}
\begin{aligned}
   R &= - \frac{2(F-1)}{w^2Y} \\
   P &= \frac{2(F-1)}{wY} \sum_i\mu^2_i d\phi_i\nn
\end{aligned}
\end{equation}
In this form one immediately notes the similarity with the nine-dimensional
Sasaki-Einstein metrics in
\cite{Cvetic:2005ft}. Note also that although to derive this form we
rescaled by $\lambda$, in the final expressions only $\lambda^2$
appears.

For the $AdS_2$ solutions we take $\lambda^2=-1$, with $w+q_i>0$ (which
implies $Y>0$ and $F<1$ so $R>0$) and $F\geq 0$. The metric on the
internal manifold $Y_9$ can then be written as
\bea
ds^2(Y_9)&=&(dz+P)^2+e^{-3A}ds^2_8 \\
&=&Fdz^2
+\frac{1-F}{w^2F}dw^2+\frac{4(1-F)}{Yw^2}
\sum_i(w+q_i)\bigg[d\mu^2_i
   +\mu^2_i\Big(d\phi_i-\frac{wdz}{2(w+q_i)}\Big)^2\bigg]\nonumber
\eea
Again this is clearly positive definite. In analogy to the type IIB
solution we choose $q_1<q_2<q_3<q_4$ such that the quartic
$\prod_i(w+q_i)-w^2$ has four roots $w_1<w_2<w_3<w_4$ and require
$w_2\leq w\leq w_3$ with $w_3>-q_1$.

The regularity conditions follow in analogy with the IIB solution,
though now the principle orbits are $U(1)^5$. For regularity at
$\mu_i=0$ one is required to take $\phi_i$ to have period $2\pi$. The
vanishing norm Killing vectors at $w=w_2$ and $w=w_3$ are given by
\begin{equation}
   l_i = c_i \del_z + c_i \sum_j \frac{w_i}{2w_i+2q_j}\del_{\phi_j} ,
\end{equation}
for $i=2,3$ and some constant $c_i$. The requirement of regularity of
the metric at these points then imposes as before
\begin{equation}
   c_i^{-1} = \sum_j \frac{w_i}{2w_i+2q_j} - 1 .
\end{equation}

We now have six different degenerations: the four points $\mu_i=0$
together with $w=w_1$ and $w=w_2$ but only five isometries. Hence we
require a relation of the form
\begin{equation}
   pl_1+ql_2+\sum_j r_j\partial_{\phi_j}=0
\end{equation}
for some co-prime integers $(p,q,r_i)$. This then further restricts
that parameters $q_i$. Since we can have $U(1)^2$ degenerations when
$w=w_i$ and $\mu_j=0$, we also require $p$ and $q$ are separately
coprime to each of the $r_i$.

To ensure that we have a good solution of M-theory we should
also ensure that the four-form is suitably quantised. We will leave
a detailed analysis of this issue for future work.


\section{Product of Kahler-Einstein Spaces}


We now turn to a different construction of K\"ahler metrics
$ds^2_{2n+2}$ satisfying~\reef{mm}. We will simply assume that it is
locally the product of a set of two-dimensional K\"ahler--Einstein
metrics
\be\label{prodans}
   ds^2_{2n+2} =\sum_{i=1}^{n+1} ds^2(KE^{(i)}_2)
\ee
where $ds^2(KE^{(i)}_2)$ is a two-dimensional K\"ahler-Einstein metric,
i.e. locally proportional to the standard metric on $S^2$, $T^2$ or
$H^2$. The Ricci form of $ds^2_{2n+2}$ is given by
\be\label{prodans2}
{\cal R}=\sum_{i=1}^{n+1}l_i J_i
\ee
where $J_i$ are the K\"ahler forms of the $ds^2(KE_2^{(i)})$
metrics and $l_i$ is zero, positive or negative depending on whether
the metric is locally that on $T^2$, $S^2$ or $H^2$, respectively.
We also have $P=\sum_i P_i$ with $dP_i= l_i J_i$ (no sum on $i$).

Globally, we will usually assume that $ds^2_{2n+2}$ extends to the
metric on a space $M_{2n+2}$ which is simply a product of
two-dimensional K\"ahler--Einstein spaces $M_{2n+2}=KE^{(1)}_2\times\dots
\times KE^{(n+1)}_2$. In the corresponding type IIB solutions
($n=2$) and $D=11$ solutions ($n=3$), one finds that the
Killing spinors are independent of the coordinates on the
$KE_2^{(i)}$. This means that the spaces $KE^{(i)}_2$ can be globally
taken to be $S^2$, $T^2$, $H^2$ or a quotient $H^2/\Gamma$, the last
giving a compact Riemann surface with genus greater than 1, while
still preserving supersymmetry.

Note that in the special case that two of the $l_i$ are equal, say
$l_1=l_2$, the analysis also covers the case when the product
$KE_2^{(1)}\times KE_2^{(2)}$ is replaced with a more general
four-dimensional K\"ahler-Einstein manifold, $KE_4$. Similar
generalisations are possible if more of the $l_i$ are equal.

Finally, in order to solve
equation~(\ref{mm}) we must impose
\be\label{cons}
\sum_{i=1}^{n+1} l_i^2 = \left(\sum_{i=1}^{n+1} l_i\right)^2
\ee
We also note that the Ricci scalar is given by
\be
R=2\sum_{i=1}^{n+1}l_i
\ee

\subsection{Type IIB $AdS_3$ solutions}
\label{3KEs}

For the type IIB $AdS_3$ case we have $n=2$.
The warp factor is
\be
e^{-4A} = \tfrac{1}{8}R = \tfrac{1}{4} \left( l_1 + l_2 + l_3 \right)
\ee
and the two form $F$ which determines the five-form flux
via~\reef{iibflux1} is given by
\be\label{exprz}
F=\frac{1}{2(l_1+l_2+l_3)}\left[
   J_1(l_2+l_3)+J_2(l_1+l_3)+J_3(l_1+l_2) \right]
\ee
The constraint \reef{cons} reads
\be
\label{rel1}
l_1l_2+l_1l_3+l_2l_3=0
\ee
and we impose $R>0$ to ensure that the warp factor is positive.

Let us analyse these constraints in more detail. Since we can permute
the spaces $KE^i_2$ we first order the parameters $l_1\leq l_2\leq
l_3$. We then observe that a rescaling of the six-dimensional K\"ahler
base space gives rise to the same $D=10$ solution (up to rescaling of
the overall factor $L$ in the ten-dimensional metric). Since
$R=2(l_1+l_2+l_3)>0$, we must have $l_3>0$ and hence we then rescale
the metric $ds^2_6$ so that $l_3=1$. Solving~\eqref{rel1} then gives
$l_2=-l_1/(l_1+1)$. Requiring $l_1\leq l_2\leq l_3$ gives a one
parameter family of solutions specified by
\be
(l_1,l_2,l_3)=(l_1,-\frac{l_1}{l_1+1},1)
\ee
with $l_1\in [-1/2,0]$.

\subsubsection*{Two equal $l_i$:}

It is interesting to look for special cases when two of the $l_i$ are
equal. As mentioned earlier, in this case we can generalise the
solution by replacing the two identical $KE_2$ factors by $KE_4$. We
find two cases. The first is when $(l_1,l_2,l_3)=(0,0,1)$ which gives
$M_6=T^4\times S^2$. This leads to the well known $AdS_3\times
S^3\times T^4$ solution corresponding to the near horizon geometry of two
intersecting D3-branes.

The second and more interesting case is when
$(l_1,l_2,l_3)=(-1/2,1,1)$ which gives $M_6=H^2\times KE_4^+$, where
$KE_4^+$ is a positively curved K\"ahler-Einstein manifold. This means
$KE^+_4$ is $S^2\times S^2$, $\bbC P^2$ or a del Pezzo $dP_k$,
$k=3,\dots,8$. It is convenient to rescale the metric so that the
$H^2$ factor has $l_1=-1$ and hence $(l_1,l_2,l_3)=(-1,2,2)$. In the
special case that $KE_4^+=\bbC P^2$, this is a solution first found by
Naka that describe D3-branes wrapping a holomorphic $H^2$
in a Calabi-Yau four-fold~\cite{naka}.

The more general solutions with arbitrary $KE_4^+$ were first given
in~\cite{gmmw3}. Let us start by rewriting the solution in a standard
form. Rescaling the metric $ds^2(KE_4^+)$ by a factor of three so that
$\mathcal{R}=6J_{KE}$, the $D=10$ solution then takes the form
\begin{equation}
\label{rescale}
\begin{aligned}
   \tfrac{\sqrt{3}}{2}ds^2 &=
      ds^2(AdS_3) + \tfrac{3}{4}ds^2(H^2) + \tfrac{9}{4}\left[
         ds^2(KE_4^+) + \tfrac{1}{9}(dz+P)^2 \right] \\
   \tfrac{3}{4}F &= \left(-\tfrac{1}{4}\right)\left[
      -\tfrac{3}{2}J_{KE}-2\vol(H_2) \right] .
\end{aligned}
\end{equation}
The term in brackets in the first line is precisely the metric on a
Sasaki--Einstein manifold, fibered over $H^2$ and with conventional
normalization factors.
To make the comparison with~\cite{gmmw3} we first note that the
conventions for the flux differ by a factor of $-1/4$. Setting
$L^2=2/\sqrt{3}$ in~\eqref{firsteq} and~\eqref{iibflux1}, and rescaling
$ds^2(KE_4^+)$ again so that $\mathcal{R}=J_{KE}$, we see
that~\eqref{rescale} is then exactly\footnote{There is a difference in
  the sign of term proportional to $J_{KE}$ in the flux, but this
  corresponds to redefining $J_{KE}\to-J_{KE}$.}  the same as that in
section 6.1 of~\cite{gmmw3} (with $d_3=0$).
It was observed in ~\cite{gmmw3} that one also obtains globally well defined solutions
for (at least some) Sasaki-Einstein manifolds in the quasi-regular class, for
which $KE_4^+$ is an orbifold.

\subsection{Type IIB bubbles}
For the corresponding type IIB bubble solutions the warp factor is given by
\be
e^{-4A} = -\tfrac{1}{8}R = -\tfrac{1}{4} \left( l_1 + l_2 + l_3 \right)
\ee
and the expression for the two-form determining the five-form flux via \reef{bubf}
is as in \reef{exprz}.
For this case we need to impose \reef{rel1} with $R<0$ instead of $R>0$.
We find a one parameter family of solutions specified by
\be
(l_1,l_2,l_3)=(-1,l_1,\frac{l_1}{l_1-1})
\ee
with $l_1\in [0,1/2]$.

It is again interesting to look for special cases when two of the $l_i$ are
equal. We find two cases. The first is when $(l_1,l_2,l_3)=(-1,0,0)$
which gives $M_6=T^4\times H^2$. The second and more interesting case
is when  $(l_1,l_2,l_3)=(-1,-1,\frac{1}{2})$ which gives $M_6=S^2\times
KE_4^-$. For example one could take $KE_4^-$ to be the four-dimesnional Bergmann metric.


\subsection{$D=11$ $AdS_2$ solutions}

For the $D=11$ case we have $n=3$. For $AdS_2$ solutions
the warp factor is given by
\be
e^{-3A} = \tfrac{1}{2}R = l_1 + l_2 + l_3 + l_4
\ee
with $R>0$.
The two-form that determines the four-form flux
via~\reef{11flux1} is given by
\be\label{exprz2}
F=-\frac{J_1(l_2+l_3+l_4)+J_2(l_1+l_3+l_4)
     +J_3(l_1+l_2+l_4)+J_4(l_1+l_2+l_3)}{l_1+l_2+l_3+l_4}
\ee

Assuming $l_1\leq l_2\leq l_3 \leq
l_4$, $R>0$ implies that $l_4\geq 0$ and hence by rescaling we can take
$l_4=1$.
We find that there is now a two parameter family of solutions labeled
by $l_1,l_2$ with
\be
l_3=-\frac{l_1l_2+l_1+l_2}{l_1+l_2+1} ,
\ee
where the ranges of $l_1$ and $l_2$ are determined by the inequalities
$l_1\leq l_2\leq l_3\leq 1$.

\subsubsection*{Three equal $l_i$:}

First consider the case when three of the $l_i$ are equal. One
immediately finds that there are a just two possibilities: The first
is when $(l_1,l_2,l_3,l_4)=(0,0,0,1)$ corresponding to $M_8=T^6\times
S^2$. This gives the well-known $AdS_2\times S^3\times T^6$ solution
of $D=11$ supergravity that arises as the near horizon limit of two
intersecting membranes.

The second is when $(l_1,l_2,l_3,l_4)=(-1,1,1,1)$ corresponding to
$M_8=KE_6^+\times H_2$. In the special case that we take $KE_6^+$ to
be a $\CP^3$ we recover the solution of $D=11$ supergravity
corresponding to the near horizon limit of a membrane wrapping a
holomorphic $H_2$ embedded in a Calabi-Yau five-fold~\cite{gkwp}.
The existence of the more general solutions for arbitrary $KE_6^+$ was
noted in a footnote in~\cite{gmmw3}.
To put the solution in a standard form, we normalise the
metric on $KE_6^+$ so that it has $\mathcal{R}=8J_{KE}$
so that the $D=11$ solution then takes the form
\bea\label{gensuk}
2^{2/3}ds^2 &=&ds^2(AdS_2)+2ds^2(H^2)
   +16\left[ds^2(KE_6^+)+\tfrac{1}{16}(dz+P)^2\right]\nn
2F &=& -\left[ 8J_{KE} + 3\vol(H^2) \right] .
\eea
Note that this has the same form as~\eqref{rescale}, with the terms in
brackets in the first line giving a Sasaki--Einstein metric, fibered
over $H^2$. In the special case that $KE_6^+=\CP^3$ this agrees with
the solution in~\cite{gkwp}.

\subsubsection*{Two equal $l_i$:}

We next consider the case when two of the $l_i$ are equal. It is
easier to take $l_1=l_2$ and $l_3\leq l_4$ instead of $l_1\leq l_2\leq
l_3\leq l_4$. We then note that if $l_1=l_2=0$ we have $l_3=0$ and hence
we have the $T^6\times S^2$ solution discussed above. Otherwise we can
always rescale so $l_1=l_2=\pm1$. This leads to two one-parameter
families of solutions
\begin{equation}
\begin{aligned}
   (l_1,l_2,l_3,l_4) &=(-1,-1, l_3, \frac{2l_3-1}{l_3-2})
      &&\qquad l_3\in(2,2+\sqrt{3}] \\
   (l_1,l_2,l_3,l_4) &= (1,1, l_3, -\frac{2l_3+1}{l_3+2})
      &&\qquad l_3\in(-2,-2+\sqrt{3}] .
\end{aligned}
\end{equation}
Note that these also contain the interesting solution
$(l_1,l_2,l_3,l_4)=(-1,-1,2+{\sqrt 3},2+{\sqrt 3})$ corresponding to
$M_8=KE_4^-\times KE_4^+$.

\subsection{$D=11$ Bubbles}

For the corresponding $D=11$ bubble solutions the warp factor is given by
\be
e^{-3A} = -\tfrac{1}{2}R = -(l_1 + l_2 + l_3 + l_4)
\ee
and the expression for the two-form determining the five-form flux via \reef{11flux1s2}
is as in \reef{exprz2}.

If we now impose $R<0$
instead of $R>0$, we find a two parameter family of solutions
specified by
\begin{equation}
   (l_1,l_2,l_3,l_4) = (-1,l_2,l_3,-\frac{l_1l_2-l_1-l_2}{l_1+l_2-1})
\end{equation}
with $-1\leq l_2\leq l_3\leq l_4$.

There are then two possibilities when three of the $l_i$ are equal:
the first is when $(l_1,l_2,l_3,l_4)=(0,0,0,-1)$ corresponding to
$M_8=T^6\times H^2$; the second is when $(l_1,l_2,l_3,l_4)=
(-1,-1,-1,1)$ corresponding to $M_8=KE_6^-\times S^2$.

We next consider the case when two of the $l_i$ are equal. We now find
the one parameter family of solutions :
\bea
(l_1,l_1,-\frac{l_1(l_1+2)}{2 l_1+1},1),\qquad l_1\in [-2-{\sqrt 3},-1]\nn
(l_1,l_1,-\frac{l_1(l_1-2)}{2 l_1-1},-1),\qquad l_1\in (-1,2-{\sqrt 3})
\eea
In addition to the cases when three $l_i$ are equal that we have
already discussed, this family also contains the interesting solution
$(-2-{\sqrt 3},-2-{\sqrt 3},1,1)$ corresponding to $M_8=KE_4^-\times
KE_4^+$.

\section{Conclusion}

It is remarkable that the equations for a generic supersymmetric
warped $AdS_3\times Y_7$ solution with $F_5$ flux in type IIB and for
a generic supersymmetric warped $AdS_2\times Y_9$ solution with
electric flux in $D=11$ supergravity are essentially the same~\cite{nak1,nak2}.
In each case the flux and local geometry of $Y_{2n+3}$ is fixed by choosing
a K\"ahler metric $ds_{2n+2}^2$ satisfying~\eqref{mm}.

Such backgrounds can arise from the near-horizon
back-reacted geometry around D3- or M2-branes wrapped on a
supersymmetric two-cycle. It is interesting to contrast these solutions
with the $AdS_5\times SE_5$ and $AdS_4\times SE_7$ solutions, where
$SE_{2n+1}$ is a Sasaki--Einstein manifold, and which arise from
unwrapped branes sitting at the apex of Ricci-flat K\"ahler cones.
Again, locally,
$SE_{2n+1}$ is determined by a choice of K\"ahler metric
$d\tilde{s}^2_{2n}$ which in this case is required to be
Einstein. From this perspective, the construction of the wrapped brane
solutions is very similar, except that the second-order tensorial
Einstein condition is replaced by the fourth-order scalar
condition~\eqref{mm} (together of course with flux which is fixed by
$ds^2_{2n+2}$).

It was pointed out in \cite{nak1,nak2} that K\"ahler metrics
satisfying \reef{mm} can also be used to construct supersymmetric
bubble solutions. In this paper we have discussed three constructions
of such K\"ahler metrics that give rise to new $AdS$ and bubble
solutions. The first construction, discussed in sections 3 and 4, is
inspired by the construction of Sasaki--Einstein metrics
in~\cite{Gauntlett:2004yd,Gauntlett:2004hh}. In this case, the
condition~\eqref{mm} could be integrated once, leaving a third-order
nonlinear differential equation~\eqref{ans} for a single function
$U(x)$. By restricting $U(x)$ to be polynomial, for type IIB we
reproduced the solutions given in~\cite{gmmw2}. For $D=11$
supergravity this led to a new one-parameter family of solutions. We
also found new non-compact $AdS_2$ $D=11$ solutions which can be
interpreted as the duals of three-dimensional CFTs coupled to
defects. It would be interesting to know whether or not there are interesting
non-polynomial solutions to the differential equation~\eqref{ans}.

The second construction of $AdS$ solutions that we discussed in section 5 was found by elucidating
the K\"ahler geometry underlying superstar solutions. These new $AdS$
solutions, which generalise those of the first construction,
are very analogous to the construction of SE metrics in \cite{Cvetic:2005ft}.
Recall that these SE metric give rise to toric Ricci-flat K\"ahler cones.
More generally, given that there are powerful techniques to study such
toric cones, it will be interesting to try and adapt these techniques
to study toric $AdS_3$ and $AdS_2$ solutions in the class of \cite{nak1,nak2}.

It is interesting that the $AdS_3$ solutions of~\cite{gmmw2}, that we
recovered here in section 3, were also recently found from a different
point of view in \cite{Kunduri:2006uh}. In that paper, an analysis of
a general class of supersymmetric $AdS$ black holes of minimal gauged
supergravity in $D=5$ was carried out. It seems likely that if one
extended the analysis of \cite{Kunduri:2006uh} from minimal gauged
supergravity to include two vector multiplets, that one would recover
the new $AdS_3$ solutions of section 5. Extending the speculations of
\cite{Kunduri:2006uh}, it is natural to wonder if the solutions that
we presented here in section 5 might describe the near horizon limit
of an asymptotically $AdS_5\times S^5$ black hole with horizon
$S^1\times Y_7$.

The third construction of solutions that we studied was to
assume that $ds^2_{2n+2}$ is locally a product of K\"ahler--Einstein metrics.
This simple approach also leads to a rich class of $AdS$ and bubble solutions.

In this paper we have focused on demonstrating that the metrics
in the new $AdS$ solutions are regular. It will be interesting to
study the topology of the solutions and then determine the additional
constraints on the parameters required to ensure that the fluxes are
suitably quantised. It will then be straightforward to calculate the
central charges of the dual SCFTs. Of particular interest, is to
identify the CFTs dual to the new $AdS$ solutions presented here. We
expect that it will be most fruitful to focus on the type IIB $AdS_3$
solutions. The similarities of the construction of the type IIB
$AdS_3$ solutions with $AdS_5\times SE_5$ solutions is suggestive that
the dual CFTs are also closely related.

\subsection*{Acknowledgements}
We would like to thank Oisin Mac Conahmna, Hong Lu, Nemani
Suryanarayana and Ricardo Ricci for helpful discussions. JPG would like to thank the Aspen Center for Physics where
some of this work was done.
NK would
like to thank the Institute for Mathematical Sciences at Imperial
College for hospitality. NK is supported by the Science Research
Center Program of the Korea Science and Engineering Foundation
(KOSEF) through the Center for Quantum Spacetime (CQUeST) of Sogang
University with grant number R11-2005-021, and by the Basics
Research Program of KOSEF with grant No. R01-2004-000-10651-0. JPG
is supported by an EPSRC Senior Fellowship and a Royal Society
Wolfson Award. DW is supported by the Royal Society through a
University Research Fellowship.
\appendix


\section{Type IIB $AdS_3$ from $D=11$ $AdS_2$ solutions}
\label{sec:appendix}


Consider a general $D=11$ solution of the form
\reef{11met}-\reef{11flux2} with an eight-dimensional K\"ahler metric
$ds^2(KE_8)$ which is locally the product of a six-dimensional K\"ahler
metric and the flat metric on a torus
\begin{equation}
   ds^2(KE_8) = ds^2(KE_6) + ds^2(T^2)
\end{equation}
Assuming that globally in $Y_9$ the flat metric extends to the
metric on a torus $T^2$, we can then dimensionally reduce to type IIA
and then T-dualise to obtain a type IIB solution. Using the formulae
in, for example, appendix~C of~\cite{gmmw3}, we deduce that the type
IIB metric is given by
\be
ds^2=e^{3A/2}\left[ds^2(AdS_2)+(dx+A_1)^2+(dz+P)^2+e^{-3A}ds^2(M_6)\right]
\ee
with $dA_1=-\vol(AdS_2)$. The first two-terms in the brackets are
simply four times the metric on a unit radius $AdS_3$ and so after
defining $e^{3A}=\tfrac{1}{4}e^{4A'}$ we can write this in the form
\be
ds^2=2e^{2A'}\left[ds^2(AdS_3)+\tfrac{1}{4}(dz+P)^2+e^{-4A'}ds^2_6 \right]
\ee
This is exactly the form of~\reef{iibmet} provided $L^2=2$.
Similarly, using the conventions of~\cite{gmmw3}, the five-form flux
can be calculated and we find
\be
-\tfrac{1}{4}F'_5 =
  4(1+*)\vol(AdS_3)\wedge\left[
     -\tfrac{1}{8}(e^{4A'}{\cal R}-4J)
     -\tfrac{1}{8}d(e^{4A'})\wedge (dz+P)\right]
\ee
which, given $L^2=2$, agrees with~\reef{iibflux1} and~\reef{iibflux2}
up to an overall difference in convention $F_5=-\frac{1}{4}F_5'$.


\begin{thebibliography}{99}

\bibitem{nak1}
  N.~Kim,
  ``AdS(3) solutions of IIB supergravity from D3-branes,''
  JHEP {\bf 0601} (2006) 094
  [arXiv:hep-th/0511029].

\bibitem{nak2}
  N.~Kim and J.~D.~Park,
  ``Comments on AdS(2) solutions of D = 11 supergravity,''
  JHEP {\bf 0609} (2006) 041
  [arXiv:hep-th/0607093] .

\bibitem{llm}
  H.~Lin, O.~Lunin and J.~M.~Maldacena,
  ``Bubbling AdS space and 1/2 BPS geometries,''
  JHEP {\bf 0410} (2004) 025
  [arXiv:hep-th/0409174].

\bibitem{donos1}
  A.~Donos,
  ``A description of 1/4 BPS configurations in minimal type IIB SUGRA,''
  arXiv:hep-th/0606199.

\bibitem{donos2}
  A.~Donos,
  ``BPS states in type IIB SUGRA with SO(4) x SO(2)(gauged) symmetry,''
  arXiv:hep-th/0610259.

\bibitem{gava}
  E.~Gava, G.~Milanesi, K.~S.~Narain and M.~O'Loughlin,
  ``1/8 BPS States in Ads/CFT,''
  arXiv:hep-th/0611065.

\bibitem{eo}
  O.~A.~P.~Conamhna and E.~O.~Colgain,
  ``Supersymmetric wrapped membranes, AdS(2) spaces, and bubbling geometries,''
  arXiv:hep-th/0612196.

\bibitem{Gauntlett:2004yd}
  J.~P.~Gauntlett, D.~Martelli, J.~Sparks and D.~Waldram,
  ``Sasaki-Einstein metrics on S(2) x S(3),''
  Adv.\ Theor.\ Math.\ Phys.\  {\bf 8} (2004) 711
  [arXiv:hep-th/0403002].


\bibitem{Gauntlett:2004hh}
  J.~P.~Gauntlett, D.~Martelli, J.~F.~Sparks and D.~Waldram,
  ``A new infinite class of Sasaki-Einstein manifolds,''
  Adv.\ Theor.\ Math.\ Phys.\  {\bf 8} (2006) 987
  [arXiv:hep-th/0403038].

\bibitem{Page:1985bq}
  D.~N.~Page and C.~N.~Pope,
  ``Inhomogeneous Einstein Metrics On Complex Line Bundles,''
  Class.\ Quant.\ Grav.\  {\bf 4} (1987) 213.


\bibitem{gmmw2}
    J.~P.~Gauntlett, O.~A.~P.~Mac Conamhna, T.~Mateos and D.~Waldram,
  ``Supersymmetric AdS(3) solutions of type IIB supergravity,''
  Phys.\ Rev.\ Lett.\  {\bf 97} (2006) 171601
  [arXiv:hep-th/0606221].


\bibitem{Myers:2001aq}
  R.~C.~Myers and O.~Tafjord,
  ``Superstars and giant gravitons,''
  JHEP {\bf 0111} (2001) 009
  [arXiv:hep-th/0109127].





\bibitem{Leblond:2001gn}
  F.~Leblond, R.~C.~Myers and D.~C.~Page,
  ``Superstars and giant gravitons in M-theory,''
  JHEP {\bf 0201} (2002) 026
  [arXiv:hep-th/0111178].









\bibitem{Cvetic:2005ft}
  M.~Cvetic, H.~Lu, D.~N.~Page and C.~N.~Pope,
  ``New Einstein-Sasaki spaces in five and higher dimensions,''
  Phys.\ Rev.\ Lett.\  {\bf 95} (2005) 071101
  [arXiv:hep-th/0504225].



\bibitem{Martelli:2005wy}
  D.~Martelli and J.~Sparks,
  ``Toric Sasaki-Einstein metrics on $S^2\times  S^3$,''
  Phys.\ Lett.\ B {\bf 621} (2005) 208
  [arXiv:hep-th/0505027].


\bibitem{naka}
  M.~Naka,
  ``Various wrapped branes from gauged supergravities,''
  arXiv:hep-th/0206141.


\bibitem{gkwp}
  J.~P.~Gauntlett, N.~Kim, S.~Pakis and D.~Waldram,
  ``Membranes wrapped on holomorphic curves,''
  Phys.\ Rev.\ D {\bf 65} (2002) 026003
  [arXiv:hep-th/0105250].


\bibitem{gmmw3}
    J.~P.~Gauntlett, O.~A.~P.~Mac Conamhna, T.~Mateos and D.~Waldram,
  ``New supersymmetric AdS(3) solutions,''
  Phys.\ Rev.\ D {\bf 74} (2006) 106007
  [arXiv:hep-th/0608055].



\bibitem{kr}
  A.~Karch and L.~Randall,
  ``Locally localized gravity,''
  JHEP {\bf 0105} (2001) 008
  [arXiv:hep-th/0011156].


\bibitem{st}
  K.~Skenderis and M.~Taylor,
  ``Branes in AdS and pp-wave spacetimes,''
  JHEP {\bf 0206} (2002) 025
  [arXiv:hep-th/0204054].


\bibitem{London:1995ib}
  L.~A.~J.~London,
  ``Arbitrary dimensional cosmological multi - black holes,''
  Nucl.\ Phys.\ B {\bf 434} (1995) 709.

\bibitem{Behrndt:1998ns}
  K.~Behrndt, A.~H.~Chamseddine and W.~A.~Sabra,
  ``BPS black holes in N = 2 five dimensional AdS supergravity,''
  Phys.\ Lett.\ B {\bf 442} (1998) 97
  [arXiv:hep-th/9807187].


\bibitem{Cvetic:1999xp}
  M.~Cvetic {\it et al.},
  ``Embedding AdS black holes in ten and eleven dimensions,''
  Nucl.\ Phys.\ B {\bf 558} (1999) 96
  [arXiv:hep-th/9903214].

\bibitem{Buchel:2006gb}
  A.~Buchel and J.~T.~Liu,
  ``Gauged supergravity from type IIB string theory on Y(p,q) manifolds,''
  arXiv:hep-th/0608002.




\bibitem{Romans:1991nq}
  L.~J.~Romans,
  ``Supersymmetric, cold and lukewarm black holes in cosmological
  Einstein-Maxwell theory,''
  Nucl.\ Phys.\ B {\bf 383} (1992) 395
  [arXiv:hep-th/9203018].


\bibitem{Gauntlett:2006ai}
  J.~P.~Gauntlett, E.~O Colgain and O.~Varela,
  ``Properties of some conformal field theories with M-theory duals,''
  arXiv:hep-th/0611219.







\bibitem{Kunduri:2006uh}
  H.~K.~Kunduri, J.~Lucietti and H.~S.~Reall,
  ``Do supersymmetric anti-de Sitter black rings exist?,''
  arXiv:hep-th/0611351.




\end{thebibliography}
\end{document}